\documentclass[revsymb,10pt,prb,twocolumn]{revtex4}%
\usepackage{graphicx}
\usepackage{amsmath}
\usepackage{amsfonts}
\usepackage{amssymb}%
\providecommand{\U}[1]{\protect\rule{.1in}{.1in}}
\newcommand{\be}{\begin{equation}}
\newcommand{\en}{\end{equation}}
\begin{document}
\title{Magnetic soliton transport over topological spin texture in chiral helimagnet
with strong easy-plane anisotropy}
\author{A.B. Borisov,$^{1}$ Jun-ichiro Kishine,$^{2}$ I.G.Bostrem,$^{3}$ and A.S.
Ovchinnikov$^{3}$}

\affiliation{$^{1}$Institute of Metal Physics, Ural Division, Russian Academy of Sciences,
Ekaterinburg 620219, Russia }

\affiliation{$^{2}$Department of Basic Sciences, Kyushu Institute of Technology, Kitakyushu
804-8550, Japan}

\affiliation{$^{3}$Department of Physics, Ural State University, Ekaterinburg, 620083 Russia}

\date{\today }

\begin{abstract}
We show the existence of an isolated soliton excitation over the topological
ground state configuration in chiral helimagnet with the Dzyaloshinskii-Moryia
exchange and the strong easy-plane anisotropy. The magnetic field
perpendicular to the helical axis stabilizes the kink crystal state which
plays a role of "topological protectorate" for the traveling soliton with a
definite handedness. To find new soliton solution, we use the B\"{a}cklund
transformation technique. It is pointed out that the traveling soliton
carries the magnon density and a magnetic solition transport may be realized.

\end{abstract}

\pacs{Valid PACS appear here}
\maketitle

\address{Institute of Metal Physics, 620219, Ekaterinburg, Russia}

\address{Faculty of Engineering, Kyushu Institute of Technology, Kitakyushu 804-8550, Japan}

\address{Department of Physics, Ural State University, 620083, Ekaterinburg, Russia}

\section{Introduction}

A study of  spatially localized excitations over non-trivial many body
"vacuum" configuration is one of the most challenging problem in condensed
matter physics. Of particular interest is a collective transport of observable
quantities accompanied with a sliding motion of incommensurate (IC) phase
modulation of charge and magnetic degrees of freedom. However, well-known
types of sliding density waves such as the charge-density wave and the collinear
spin-density-wave cannot easily be observed because the internal phase
modulation does not carry directly measured quantity.\cite{PWABasicNotions}
Another example is the charge and spin soliton transport in conjugated polymers
where the double degeneracy of  ground state configurations gives rise to
diffusive solitons.\cite{Fukuyama-Takayama} In this paper, we demonstrate a
new possibility to create an isolated magnetic soliton over the topological
ground state configuration in chiral helimagnet with the antisymmetric
Dzyaloshinskii-Moryia (DM) exchange and the strong easy-plane anisotropy. We show that  the 
isolated magnetic soliton exists and  it  can carry the observable magnetic density.

Helimagnetic structures are stabilized by  either frustration among exchange
couplings\cite{Yoshimori59} or the DM relativistic
exchange.\cite{Dzyaloshinskii58} 
In the latter case, an absence of the rotoinversion symmetry in chiral crystals
causes the Lifshitz invariant in the Landau free energy. Consequently, a 
long-period incommensurate helimagnetic structure is stabilized with the
definite (left-handed or right-handed) chirality fixed by the direction of the
DM vector. Over the past two decades, the ground state and linear excitations
of the chiral helimagnet under an applied magnetic field have been a subject of extensive
studies from both experimental and theoretical viewpoints.\cite{Izyumov}
Concerning nonlinear excitations in this class of spiral systems, the solitonic
excitations were
studied in the case of \textit{easy-axis} anisotropy.\cite{Borisov} It has been revealed that the first simplest soliton solution
in the easy-axis helimagnet presents a helical domain wall, a nucleation of a
helical phase, whereas the second solution, so-called a wave of rotation,
describes a localized change of the phase with a finite velocity, which is
accompanied by a coming of moments out of the basal plane (Fig. 1). It turns
out that the energy of the wave of rotation is less than that of spin wave
with the same linear momentum. In a presence of magnetic field applied along
the easy-axis, when a simple helimagnetic structure transforms into the
conical one, there are again two solitons with energies less then the energy
of spin wave with the same momentum.%
\begin{figure}
[ptb]
\begin{center}
\includegraphics[
height=2.7985in,
width=2.0773in
]%
{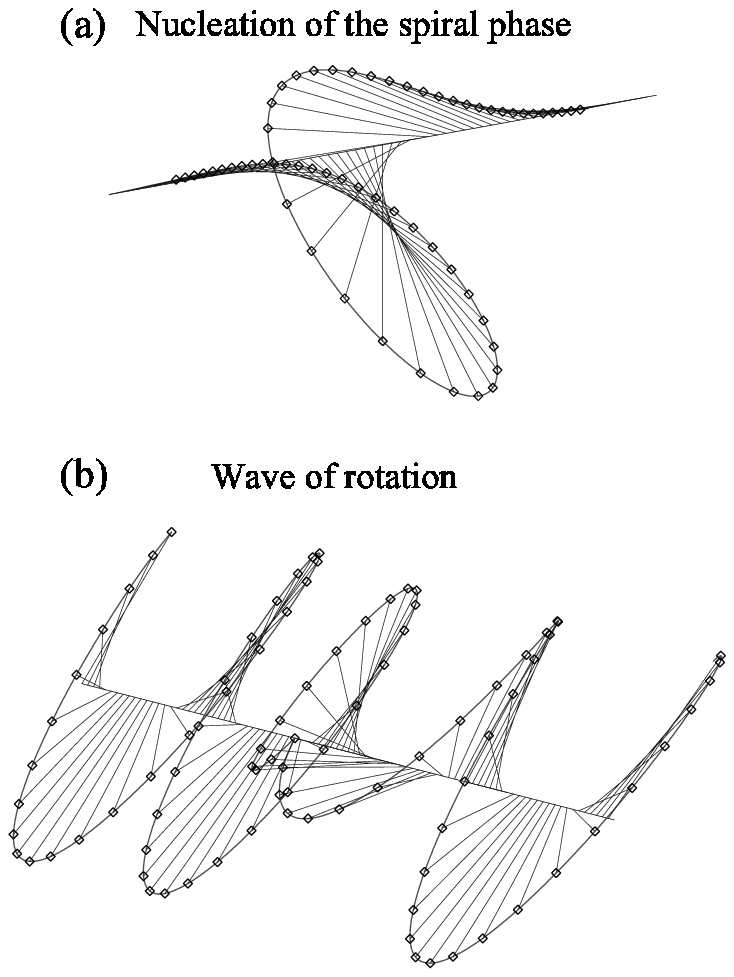}%
\caption{Nonlinear excitations in a helimagnet with an easy axis
anisotropy: the nucleation of the spiral phase (a) and the localized wave of
rotation (b).}%
\end{center}
\end{figure}

Recent progress on synthesis of new class of helimagnetic structures revives
the interest to the case of chiral helimagnet with an easy-plane
anisotropy.\cite{KIY} Findings of solitons in the previous (easy-axis) case
was based on a deep analogy between a dynamics of these nonlinear excitations
spreading over the easy axis in the chiral helimagnet and a dynamics of
nonlinear excitations in the easy-axis ferromagnet. The correspondence reached
by a gauge transformation enables to use a well established
classification of nonlinear excitations in the last system.\cite{Sklyanin}
However, the theoretical tool turns out to be inappropriate to be applied to
the helimagnet with the Lifshitz invariant an the easy-plane anisotropy
because the axial symmetry is lost in this case.

In this paper, we show that this problem is resolved with the B\"{a}cklund
transformation (BT) technique.\cite{Rogers} The method we use has been
effectively applied to studies of topological vortex-type singular solutions
of the elliptic sine-Gordon (SG) equation.\cite{Leibbrandt} In particular,
one- and two-dimensional vortex lattices on both a homogeneous and periodic
backgrounds have been constructed using the B\"{a}cklund
transformation.\cite{Borisov88} One of the main findings of our investigation
is an appearance of a non-trivial traveling soliton which carries a localized
magnon density. This result may be useful in spintronics
technology.\cite{Sloncew,Berger} {In Sec. II, we describe the model
Hamiltonian and basic equations. In Sec. III, we present the method to derive
the novel soliton solution by using the }B\"{a}cklund transformations. {In
Sec. IV, we discuss the energy and momentum associated with a creation of the
soliton. In Sec. V, we demonstrate that the traveling soliton carries the
magnon density. Finally, we summarize the results in Sec. VI. }

\section{Basic equations}

We describe the chiral helimagnet by the continuum Hamiltonian density,%
\begin{align}
\mathcal{H}  &  =\frac{\alpha}{2}\left(  \frac{\partial\mathbf{M}}{\partial
z}\right)  ^{2}+D\left(  M_{x}\frac{\partial M_{y}}{\partial z}-M_{y}%
\frac{\partial M_{x}}{\partial z}\right) \nonumber\\
&  +\beta^{2}M_{z}^{2}+h_{x}M_{x}, \label{Ham}%
\end{align}
where the symmetric exchange coupling strength is given by $\alpha>0$, the
mono-axial DM coupling strength is given by $D,$ the easy-plane anisotropy
strength is given by $\beta^{2}>0$, and the external magnetic field applied
perpendicular to the chiral $z$ axis is $\mathbf{h}=(h_{x},0,0).$ {When the 
model Hamiltonian (\ref{Ham}) is written, one implicitly assumes that the
magnetic atoms form a cubic lattice and the uniform ferromagnetic coupling
exists between the adjacent chains to stabilize the long-range order. Then,
the Hamiltonian is interpreted as an effective  one-dimensional model based on the
interchain mean field picture. }

Starting with (\ref{Ham}), one derives the Landau-Lifshitz equation
$\partial\mathbf{M}/\partial t=\left[  \mathbf{M}\times\mathbf{H}%
_{eff}\right]  $, where $H_{eff}^{i}=\delta\mathcal{H}/\delta M_{i}$
($i=x,y,z$) is the effective field acting on the magnetic moment $\mathbf{M}$,
the constant $\alpha$ is a feature of the symmetrical exchange coupling. In
the explicit form these equations read as
\begin{equation}
\left\{
\begin{array}
[c]{c}%
\dfrac{\partial M_{x}}{\partial t}=2\beta^{2}M_{y}M_{z}+2DM_{z}\dfrac{\partial
M_{x}}{\partial z}\\
+\alpha M_{z}\dfrac{\partial^{2}M_{y}}{\partial z^{2}}-\alpha M_{y}%
\dfrac{\partial^{2}M_{z}}{\partial z^{2}},\\
\dfrac{\partial M_{y}}{\partial t}=-2\beta^{2}M_{x}M_{z}+h_{x}M_{z}%
+2DM_{z}\dfrac{\partial M_{y}}{\partial z}\\
-\alpha M_{z}\dfrac{\partial^{2}M_{x}}{\partial z^{2}}+\alpha M_{x}%
\dfrac{\partial^{2}M_{z}}{\partial z^{2}},\\
\dfrac{\partial M_{z}}{\partial t}=-h_{x}M_{y}-2DM_{y}\dfrac{\partial M_{y}%
}{\partial z}-2DM_{x}\dfrac{\partial M_{x}}{\partial z}\\
+\alpha M_{y}\dfrac{\partial^{2}M_{x}}{\partial z^{2}}-\alpha M_{x}%
\dfrac{\partial^{2}M_{y}}{\partial z^{2}}.
\end{array}
\right.  \label{LLEq}%
\end{equation}
By using the polar angles $\theta(z)$\ and $\varphi(z)$, we represent
$\mathbf{M}(z)=\left(  \sin\theta\cos\varphi,\sin\theta\sin\varphi,\cos
\theta\right)  $ and then Eqs.(\ref{LLEq}) are transformed into
\begin{equation}
\frac{\partial\theta}{\partial t}=-b\sin\varphi-\cos\theta\left(
a+2\frac{\partial\varphi}{\partial z}\right)  \frac{\partial\theta}{\partial
z}-\sin\theta\frac{\partial^{2}\varphi}{\partial z^{2}}, \label{TPEq0}%
\end{equation}
and%
\begin{align}
\frac{\partial\varphi}{\partial t}  &  =2\beta^{2}\cos\theta-\cos\theta\left(
\frac{\partial\varphi}{\partial z}\right)  ^{2}-b\cos\varphi\cot
\theta\label{TPEq1}\\
&  -a\cos\theta\frac{\partial\varphi}{\partial z}+\frac{1}{\sin\theta}%
\frac{\partial^{2}\theta}{\partial z^{2}}.\nonumber
\end{align}
Hereinafter, we set $\alpha=1,$ and rewrite $h_{x}=b$, and $D=a/2$. The
magnetic kink crystal phase is described by the stationary soliton solution
minimizing $\mathcal{H}$ with keeping $\theta=\pi/2$. The solution obeys the
SG equation, $\varphi_{zz}=-b\sin\varphi,$ and is given by $\sin
(\varphi/2)=\text{sn}(\sqrt{b}z,q),$where $\mathrm{sn}$ is the Jacobi elliptic
function with the elliptic modulus $q$ ($0<q^{2}<1$). The magnetic kink
crystal phase is sometimes called the soliton lattice (SL) state. The
background material behind this solution is summarized in Appendix A. As the
modulus $q$ increases, the lattice period of the kink crystal increases and
finally diverges at $q\rightarrow1$, where the incommensurate-to-commensurate
(IC-C)\ phase transition occurs.

Now, our goal is to find out possible nonlinear excitations with small
deflections of spins around the basal $xy$ plane. For this purpose we use the
expansion
\begin{equation}
\theta(z,t)=\frac{\pi}{2}+\tilde{s}\theta_{1}(z,t) \label{ExpanTh}%
\end{equation}
with the small fluctuation $\theta_{1}\ll1$ with $\tilde{s}$ being a dummy
variable controlling an order of the expansion. Plugging this into
Eqs.(\ref{TPEq0},\ref{TPEq1}), we have
\begin{equation}
0=b\sin\varphi+\tilde{s}\frac{\partial\theta_{1}}{\partial t}+\frac
{\partial^{2}\varphi}{\partial z^{2}}, \label{sexp2}%
\end{equation}
and%
\begin{equation}
\frac{\partial\varphi}{\partial t}+\tilde{s}\theta_{1}\left[  2\beta
^{2}-\left(  \frac{\partial\varphi}{\partial z}\right)  ^{2}-b\cos
\varphi-a\frac{\partial\varphi}{\partial z}\right]  -\tilde{s}\frac
{\partial^{2}\theta_{1}}{\partial z^{2}}=0, \label{sexp1}%
\end{equation}
where the terms linear in $\tilde{s}$ are hold.

In real materials,\cite{KIY} the order of magnitude of the DM coupling gives
the long-period wave vector $Q_{0}\approx D\sim10^{-2}$ and one obtains
$\varphi_{z}\approx Q_{0}\sim10^{-2}$. Magnetic field strength $b$ used in
experiment are of order $10^{-5}$. The further analytical treatment is
performed in a regime of the \textit{strong easy-plane anisotropy}, i.e.
$\beta^{2}\gtrsim10^{-3}.$ For spin configurations, where the last term in
Eq.(\ref{sexp1}) may be neglected (see the end of Sec. III), one obtains the relation%
\begin{equation}
\theta_{1}=-\frac{1}{2\tilde{s}\beta^{2}}\frac{\partial\varphi}{\partial t},
\label{Th1}%
\end{equation}
which establishes a conjugate relation between the dynamical $\theta$ and
$\varphi$ variables. The same relation was discussed in the context of soliton
dynamics in one-dimensional magnets.\cite{Mikeska81} Plugging Eq.(\ref{Th1})
into Eq.(\ref{sexp2}), we have the (1+1)-dimensional SG equation,%
\begin{equation}
b\sin\varphi-\frac{1}{2\beta^{2}}\frac{\partial^{2}\varphi}{\partial t^{2}%
}+\frac{\partial^{2}\varphi}{\partial z^{2}}=0 \label{SineGord}%
\end{equation}
We use Eqs.(\ref{Th1}) and (\ref{SineGord}) to find the soliton solutions.

\section{B\"{a}cklund transformation.}

The BT is known to be a powerful method to systematically construct non-linear
solution in a certain class of partial differential equations. The case of the
sine-Gordon equation is briefly summarized in Appendix B. In the new
coordinates $\sqrt{2b}\beta t\rightarrow t$, $\sqrt{b}z\rightarrow z$ , two
solutions $\varphi$ and $\tilde{\varphi}$ of the SG equation $\varphi
_{tt}-\varphi_{zz}=\sin\varphi$ are related via the BT with the real valued
parameter $k$.\cite{Dodd}%
\begin{equation}
\left\{
\begin{array}
[c]{c}%
\tilde{\varphi}_{t}=\varphi_{z}+k\sin\left(  \dfrac{\varphi+\tilde{\varphi}%
}{2}\right)  -\dfrac{1}{k}\sin\left(  \dfrac{\varphi-\tilde{\varphi}}%
{2}\right)  ,\\
\tilde{\varphi}_{z}=\varphi_{t}+k\sin\left(  \dfrac{\varphi+\tilde{\varphi}%
}{2}\right)  +\dfrac{1}{k}\sin\left(  \dfrac{\varphi-\tilde{\varphi}}%
{2}\right)  .
\end{array}
\right.  \label{BackTran}%
\end{equation}
We here use the notation $\varphi_{z}=\partial_{z}\varphi=\partial
\varphi/\partial z$, $\varphi_{zz}=\partial_{z}^{2}\varphi=\partial^{2}%
\varphi/\partial z^{2}$, and so on. A pair of the solutions are written in a
form,%
\begin{equation}
\left\{
\begin{array}
[c]{c}%
\varphi\left(  z\right)  =\pi+4\tan^{-1}\left[  F(z)\right]  ,\\
\tilde{\varphi}\left(  z,t\right)  =\pi+4\tan^{-1}\left[  V(z,t)\right]  .
\end{array}
\right.  \label{inspection}%
\end{equation}
It is easy to verify that the background kink crystal solution, $\sin\left(
\varphi/2\right)  =\text{sn}\left(  z,q\right)  ,$ is reproduced by choosing
$F(z)$\ as%
\begin{equation}
F(z)=\frac{\text{cn}\left(  z,q\right)  }{1+\text{sn}\left(  z,q\right)  }.
\label{FZ}%
\end{equation}
The dependence of $V$ upon the time and the coordinate should be found through
the BT.

To reduce a complexity, it is convenient to rewrite Eqs.(\ref{BackTran}) in
terms of the functions $F$ and $V$. Then, we have the first and the second BTs
respectively given by%
\begin{gather}
V_{t}=\frac{2kF_{z}+F(k^{2}+1)}{2k(1+F^{2})}V^{2}+\frac{(F^{2}-1)(k^{2}%
-1)}{2k(1+F^{2})}V\nonumber\\
+\frac{2kF_{z}-F(k^{2}+1)}{2k(1+F^{2})}, \label{Tdep}%
\end{gather}
and%
\begin{equation}
V_{z}=\frac{V\left(  F^{2}-1\right)  (k^{2}+1)+F(V^{2}-1)(k^{2}-1)}%
{2k(1+F^{2})}. \label{Zdep}%
\end{equation}
The further strategy is straightforward. The time dependence is firstly found
from Eq.(\ref{Tdep}) and then followed by solving of Eq.(\ref{Zdep}).

\subsection{BT Equation (\ref{Tdep})}

The right-hand side of Eq.(\ref{Tdep}) is quadratic in $V$. To reach a
simplification we use the shift
\begin{equation}
V(z,t)=U(z,t)-\frac{(F^{2}-1)(k^{2}-1)}{2\left(  Fk^{2}+2kF_{z}+F\right)  }
\label{VviaU}%
\end{equation}
that transforms Eq.(\ref{Tdep}) into
\begin{equation}
U_{t}+A(z)U^{2}+B(z)=0, \label{UAB}%
\end{equation}
where
\begin{equation}
A(z)=-\frac{Fk^{2}+2kF_{z}+F}{2k(1+F^{2})}, \label{cfA}%
\end{equation}
and%
\begin{align}
B(z)  &  =\left[  8k(1+F^{2})\left(  Fk^{2}+2kF_{z}+F\right)  \right]
^{-1}\nonumber\\
&  \times\left[  -16k^{2}F_{z}^{2}+F^{4}(k^{2}-1)^{2}\right. \nonumber\\
&  \left.  +(k^{2}-1)^{2}+2F^{2}(1+k^{4}+6k^{2})\right]  . \label{cfB}%
\end{align}
Another simplification is achieved through the identity (see Appendix C),%
\begin{equation}
A(z)B(z)=\frac{1}{16}\left(  \frac{4}{q^{2}}-k^{2}-\frac{1}{k^{2}}-2\right)
=s. \label{AB}%
\end{equation}

In the sector $s<0$, when the B\"{a}cklund parameter $k$ is constrained by
$\left\vert k+k^{-1}\right\vert >2/q$, Eq.(\ref{UAB}) can be immediately
resolved
\begin{equation}
U\left(  z,t\right)  =\frac{\mathcal{S}}{A(z)}\tanh\left[  \frac{\mathcal{S}%
}{A(z)}\left\{  A(z)t-C_{1}(z)\right\}  \right]  , \label{Ut}%
\end{equation}
where $s=-\mathcal{S}^{2}$, and $C_{1}(z)$ is a function of the coordinate.
The another sector $s>0$ contains no localized solitons. The time dependence
that we need is recovered from Eq.(\ref{VviaU})
\begin{align}
V(z,t)  &  =\frac{\mathcal{S}}{A(z)}\tanh\left[  \frac{\mathcal{S}}%
{A(z)}\left\{  A(z)t-C_{1}(z)\right\}  \right] \nonumber\\
&  -\frac{(k^{2}-1)(F^{2}-1)}{2\left(  Fk^{2}+2kF_{z}+F\right)  }. \label{Vzt}%
\end{align}

\subsection{BT Equation (\ref{Zdep})}

The unknown function $C_{1}(z)$ should be determined from the second
BT\ equation. The derivation is relegated to Appendix D and the result has the
form
\begin{equation}
\frac{A_{z}}{A}\left(  At-C_{1}\right)  -\left(  A_{z}t-C_{1z}\right)
=\frac{F(k^{2}-1)}{2k(1+F^{2})}. \label{SMPLB2}%
\end{equation}
The substitution $C_{1}(z)=A(z)\mathcal{M}(z)$ transforms this equation into
\[
\mathcal{M}_{z}(z)=\frac{F(k^{2}-1)}{2Ak(1+F^{2})}.
\]
By using the explicit expressions for $F(z)$ and
\begin{equation}
A(z)=\frac{1}{4k}\left(  \frac{2k}{q}\text{dn}\left(  z,q\right)  -\left(
k^{2}+1\right)  \text{cn}\left(  z,q\right)  \right)  , \label{Az}%
\end{equation}
one obtain
\begin{equation}
\mathcal{M}_{z}(z)=\frac{(1-k^{2})\,q\,\text{cn}\left(  z,q\right)  }%
{(1+k^{2})\,q\,\text{cn}\left(  z,q\right)  -2k\,\text{dn}\left(  z,q\right)
}. \label{Mzder}%
\end{equation}
After integration, this yields (see Appendix E)
\begin{align}
\mathcal{M}(z)  &  =\frac{1}{4q{\mathcal{S}}}\log\left\vert \frac
{4{\mathcal{S}}k-(k^{2}-1)\,\text{sn}(z,q)}{4{\mathcal{S}}k+(k^{2}%
-1)\,\text{sn}(z,q)}\right\vert -\frac{k^{2}+1}{k^{2}-1}z\nonumber\\
&  +\frac{4k^{2}(1-q^{2})(1+k^{2})}{(1-k^{2})[(1+k^{2})^{2}q^{2}-4k^{2}%
]}{\tilde{\Pi}}(n,\text{am}(z,q),q^{2}). \label{Mzsol}%
\end{align}
The function $\tilde{\Pi}$ is defined by
\begin{align*}
&  {\tilde{\Pi}}(n,\text{am}(z,q),q^{2})\\
&  =\left\{
\begin{array}
[c]{c}%
\Pi(n,\text{am}(z,q),q^{2})\quad\text{for}\quad n<1,\\
-\Pi(N,\text{am}(z,q),q^{2})+\mathcal{F}(\text{am}(z,q),q^{2})\\
+(1/2p_{1})\log\left\vert \dfrac{\text{cn}(z,q)\text{dn}(z,q)+p_{1}%
\,\text{sn}(z,q)}{\text{cn}(z,q)\text{dn}(z,q)-p_{1}\,\text{sn}(z,q)}%
\right\vert \\
\text{for}\quad n>1,
\end{array}
\right.
\end{align*}
where $\text{am}$ is the Jacobi elliptic function, and $n={(1-k^{2})^{2}q^{2}}/{[(1+k^{2})^{2}q^{2}-4k^{2}]}$ is the
characteristic index of the elliptic integral of the third kind $\Pi$. Furthermore, $N=q^{2}/n$, $p_{1}=\sqrt{(n-1)(1-N)}$, and $\mathcal{F}(...)$
is the elliptic integral of the first kind (see formulas 17.7.7 and 17.7.8 in
Ref.\cite{AS}). Eqs.(\ref{Vzt},\ref{Az},\ref{Mzsol}) determine explicitly the
solution conjugated to the soliton lattice through the B\"{a}cklund
transformation. To complete we focus on the asymptotic behavior of the found solution.

For definiteness we choose $|k|>1$, the opposite case $|k|<1$ can be
analogously treated. By noting that $\mathcal{M}(z\rightarrow\pm\infty
)=\pm\infty$ in this case, and vise versa for $|k|<1$, one obtains
\begin{align}
V(z  &  \rightarrow\pm\infty,t)\approx\mp\frac{\mathcal{S}}{A(z)}-\frac
{(k^{2}-1)(F^{2}-1)}{2\left(  Fk^{2}+2kF_{z}+F\right)  }\nonumber\\
&  =\frac{k^{2}-1\pm4\mathcal{S}k+F^{2}(1-k^{2}\pm4\mathcal{S}k)}{2\left(
Fk^{2}+2kF_{z}+F\right)  }\nonumber\\
&  =q\frac{[(k^{2}-1)\text{sn}(z)\pm4k\mathcal{S}]}{(1+k^{2})q\,\text{cn}%
(z)-2k\text{dn}(z)}, \label{Vzappr}%
\end{align}
if to use the explicit expressions for $F(z)$ and $F_{z}(z)$.

Now, it is easy to prove that Eq.(\ref{Vzappr}) may be rewritten in the form
\begin{equation}
V(z\rightarrow\pm\infty,t)\approx\frac{\text{cn}(z+\delta,q)}{1+\text{sn}%
(z+\delta,q)}, \label{Phinew}%
\end{equation}
which is similar to that used for $F(z)$ given by Eq.(\ref{FZ}). To find the
shift $\delta$ and explicit values of sn$(\delta,q)$, cn$(\delta,q)$, and
dn$(\delta,q),$ we use the addition theorems for the elliptic functions,%
\begin{align*}
\text{cn}(z+\delta)  &  =\frac{\text{cn}(z)\text{cn}(\delta)-\text{sn}%
(z)\text{sn}(\delta)\text{dn}(z)\text{dn}(\delta)}{1-q^{2}\text{sn}%
^{2}(z,q)\text{sn}^{2}(\delta,q)},\\
\text{sn}(z+\delta)  &  =\frac{\text{sn}(z)\text{cn}(\delta)\text{dn}%
(\delta)+\text{sn}(\delta)\text{dn}(z)\text{cn}(z)}{1-q^{2}\text{sn}%
^{2}(z)\text{sn}^{2}(\delta)},
\end{align*}
where the elliptic modulus $q$\ is omitted. Recalling the periodicity of the
form (\ref{Phinew}) and requiring the function given by Eq.(\ref{Phinew}) to
coincide with the values of Eq.(\ref{Vzappr}) in the boundary points $0$ and
$2K$ of the period, we obtain the desired result,%
\begin{equation}
\left.
\begin{array}
[c]{c}%
\text{sn}(\delta,q)=-\dfrac{2k}{(1+k^{2})q},\\
\text{cn}(\delta,q)=\pm\dfrac{4k\mathcal{S}}{1+k^{2}},\\
\text{dn}(\delta,q)=\dfrac{1-k^{2}}{1+k^{2}}.
\end{array}
\right.  \label{alphaEq}%
\end{equation}
From this result, we obtain the asymptotic behavior,%
\[
\tilde{\varphi}\left(  z\rightarrow\pm\infty,t\right)  \approx\pi+4\tan
^{-1}\left[  \frac{\text{cn}(z+\delta,q)}{1+\text{sn}(z+\delta,q)}\right]  ,
\]
where the sign plus in Eqs.(\ref{alphaEq}) is related with the limit
$z\rightarrow\infty$, whereas the minus is did with $z\rightarrow-\infty$. At
the end, we note an analogy of the introduced shift $\delta$ with the shift of
atoms relative to potential minima in the model of Frank and Van der Merwe
(FVdM).\cite{Bak} Now, we have done everything we need by using BT and found
out that the BT creates an additional kink over the background kink crystal
state and causes an expansion of the periodical spin structure at infinity.

\subsection{The case near the incommensurate-to-commensurate phase boundary}

We consider in details the case near the IC-C phase boundary ($q\rightarrow1$). In this
limit, the lattice period of the kink crystal tends to go to infinity and the
background state consists of a solitary kink described by $\varphi
(z)=2\sin^{-1}\left(  \tanh z\right)  $. Fortunately, in this limit, Eq.
(\ref{Mzder}) can be integrated by using elementary functions. By using the
relationships $\text{cn}\left(  z,1\right)  =1/\cosh  z  $,
$\text{sn}\left(  z,1\right)  =\tanh z $ and $\text{dn}\left(
z,1\right)  =1/\cosh z$, Eq.(\ref{Mzder}) is integrated to
give
\[
\mathcal{M}(z)=\frac{1+k}{1-k}z+\mathcal{M}_{0},
\]
where $\mathcal{M}_{0}$ is a constant. The same simplifications yield
\[
A(z)=-\frac{(k-1)^{2}}{4k\cosh z  },
\]
and $F_z (z,1)=-\left[\cosh   z  (1+\tanh  z)
\right]  ^{-1}$, $F(z,1)=\left[\cosh   z  (  1+\tanh z ) \right] ^{-1}$. The parameter $\mathcal{S}$ in Eq.(\ref{Ut}) reads
as $\mathcal{S=}\left(  k^{2}-1\right)  /\left(  4k\right)  $ in this case.

After collecting the results together, one eventually obtain
\begin{gather}
V(z,t)=\frac{k+1}{k-1}\nonumber\\
\times\left\{  \sinh z  +\cosh  z  \tanh\left[
\frac{k^{2}-1}{4k}\left(  \frac{1+k}{1-k}z+\mathcal{M}_{0}-t\right)  \right]
\right\}  ,\nonumber\\
\end{gather}
and%
\begin{equation}
\tilde{\varphi}\left(  z,t\right)  =\pi+4\tan^{-1}\left[  V(z,t)\right]  ,
\label{phiAn}%
\end{equation}
where $\upsilon_{0}=\left(  1-k\right)  /\left(  1+k\right)  $ can be thought
of as the \textquotedblright velocity\textquotedblright\ of the soliton.

The polar angle computed via Eq.(\ref{Th1}) takes the form,%
\begin{align}
\theta(z,t)  &  =\frac{\pi}{2}+\sqrt{\frac{b}{2\beta^{2}}}\frac{\left(
k+1\right)  ^{2}}{k}\frac{1}{1+V^{2}}\nonumber\\
&  \times\frac{\cosh  z  }{\cosh^{2}\left\{  \dfrac{k^{2}-1}%
{4k}\left(  \dfrac{1+k}{1-k}z+\mathcal{M}_{0}-t\right)  \right\}  }.
\label{explthet}%
\end{align}
This solution is viewed as a "collision"\ of two kinks as shown in Fig 2. One
of the kinks is "at rest" as a member of the background\ configuration, while
the other travels with the speed $\upsilon_{0}$ and go through the background
without changing its shape.

To justify the inequality $|d^{2}\theta_{1}/dz^{2}|\ll2\beta^{2}\theta_{1}$
that have been assumed to obtain Eq.(\ref{Th1}) the restriction
\begin{equation}
\left\vert k+\frac{1}{k}\right\vert \ll4\beta^{2} \label{restr_K}%
\end{equation}
must be fulfilled, which is derived by means of Eq.(\ref{explthet}). To obtain
this condition, we use the asymptotic of $|(d^{2}\theta_{1}/dz^{2})/\theta
_{1}|$ at $z\rightarrow\pm\infty$. Note that the constraint $|k+1/k|>2/q$
excludes only the points $k=\pm1$. Therefore, there is a region of finite $k$
values provided $\beta^{2}\gg1/2$.%
\begin{figure}
[ptb]
\begin{center}
\includegraphics[
height=2.7294in,
width=2.2001in
]%
{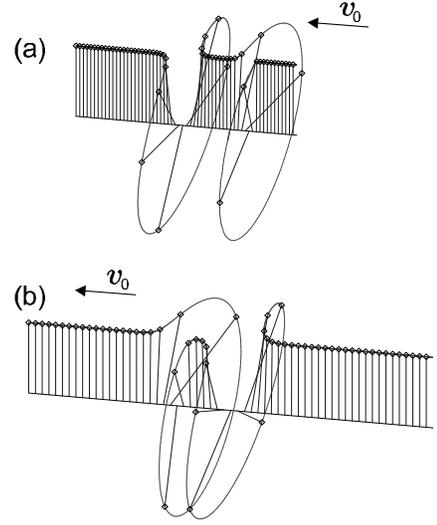}%
\caption{A profile of the soliton excitation at $q\to 1$ in a spiral with an easy
plane anisotropy directly before (a) and after (b) \textquotedblright
collision\textquotedblright\ of the kinks. }%
\end{center}
\end{figure}

\section{Energy and momentum}

Next we consider the {energy and momentum associated with creation of the
found soliton. }The energy density (\ref{Ham}) is written in polar coordinates
as
\begin{equation}
\mathcal{H}=\frac{1}{2}\left(  \theta_{z}^{2}+\sin^{2}\theta\varphi_{z}%
^{2}\right)  +\frac{a}{2}\sin^{2}\theta\varphi_{z}+\beta^{2}\cos^{2}%
\theta+b\sin\theta\cos\varphi. \label{Hpolar}%
\end{equation}
Using the expansion (\ref{ExpanTh},\ref{Th1}) and neglecting terms higher than
second-order derivatives one obtains
\begin{equation}
\mathcal{H}\approx\frac{1}{2}\varphi_{z}^{2}+\frac{a}{2}\varphi_{z}%
+\frac{\varphi_{t}^{2}}{4\beta^{2}}+b\cos\varphi\left(  1-\frac{\varphi
_{t}^{2}}{8\beta^{4}}\right)  . \label{ApproxEn}%
\end{equation}
In the case of small fields, the term proportional to $b\varphi_{t}^{2}$ may
be ignored and the energy density $\mathcal{H}$ measured in $b$ units becomes
\[
\mathcal{H}=\cos\varphi+\frac{1}{2}\left(  \varphi_{z}^{2}+\varphi_{t}%
^{2}\right)  +c\,\varphi_{z},
\]
where $c=a/\left(  2\sqrt{b}\right)  $ and the coordinates $\sqrt{2b}\beta
t\rightarrow t$, $\sqrt{b}z\rightarrow z$ are again used. Following the method
suggested in Ref.\cite{Kulagin}, we find the difference between the energy
densities calculated on the solutions coupled by the BT
\begin{gather}
\mathcal{H}(\tilde{\varphi})\mathcal{-H}(\varphi)=c\,\left(  \tilde{\varphi
}_{z}-\varphi_{z}\right) \nonumber\\
+\frac{1}{k^{2}}\left[  k^{2}\sin\left(  \frac{\varphi+\tilde{\varphi}}%
{2}\right)  +\sin\left(  \frac{\varphi-\tilde{\varphi}}{2}\right)  \right]
^{2}\nonumber\\
+\left[  k\sin\left(  \frac{\varphi+\tilde{\varphi}}{2}\right)  +\frac{1}%
{k}\sin\left(  \frac{\varphi-\tilde{\varphi}}{2}\right)  \right]  \varphi
_{t}\nonumber\\
+\left[  k\sin\left(  \frac{\varphi+\tilde{\varphi}}{2}\right)  -\frac{1}%
{k}\sin\left(  \frac{\varphi-\tilde{\varphi}}{2}\right)  \right]  \varphi_{z},
\label{Ediffer}%
\end{gather}
where the transformations (\ref{BackTran}) are employed. This amounts to a
derivative of the function
\begin{equation}
\Psi^{(e)}=c\,\left(  \tilde{\varphi}-\varphi\right)  +\frac{2}{k}\cos\left(
\frac{\varphi-\tilde{\varphi}}{2}\right)  -2k\cos\left(  \frac{\varphi
+\tilde{\varphi}}{2}\right)  \label{Fdiff}%
\end{equation}
with respect to $z$. Thus, the difference of the energies (\ref{Ediffer})
integrated over the total length $L$ of the system is equal to%
\[
\int\limits_{0}^{L}dz\,\left\{  \mathcal{H}(\tilde{\varphi})\mathcal{-H}%
(\varphi)\right\}  =\Psi^{(e)}\left(  L\right)  -\Psi^{(e)}\left(  0\right)
.
\]

The momentum density $\mathcal{P}=\hbar S\left(  1-\cos\theta\right)
\varphi_{z}$ is treated in the same manner. The spin value $S$ is realted with the magnetic moment by $\textbf{M}=2\mu_0 \textbf{S}$. Using again the expansions (\ref{ExpanTh},\ref{Th1}) and the new coordinates one obtains
\[
\mathcal{P}\approx\varphi_{z}-\nu\varphi_{z}\varphi_{t}%
\]
of the same accuracy as in the case (\ref{ApproxEn}). Here, the momentum
$\mathcal{P}$ is measured in the units $\hbar S\sqrt{b}$, and $\nu
=\sqrt{b/2\beta^{2}}$. The difference between the momentum densities of the
solutions conjugated by the BT amounts to%
\[
\mathcal{P}(\tilde{\varphi})\mathcal{-P}(\varphi)=\Psi_{z}^{(m)},
\]
where the function $\Psi^{(m)}$\ is given by%
\begin{equation}
\Psi^{(m)}=\tilde{\varphi}\mathcal{-}\varphi+\nu\left[  \frac{2}{k}\cos\left(
\frac{\varphi-\tilde{\varphi}}{2}\right)  +2k\cos\left(  \frac{\varphi
+\tilde{\varphi}}{2}\right)  \right]  . \label{Mdiff}%
\end{equation}
Therefore, the additional momentum with reference to the background kink
crystal state becomes%
\[
\int\limits_{0}^{L}dz\,\left\{  \mathcal{P}(\tilde{\varphi})\mathcal{-P}%
(\varphi)\right\}  =\Psi^{(m)}\left(  L\right)  -\Psi^{(m)}\left(  0\right)
.
\]

\section{Magnon current.}

The magnon current\ transferred by the $\theta$-fluctuations is determined
through the definition of the accumulated magnon density $\mathcal{\rho
}_{\text{s}}$ in the total magnon density $\mathcal{N}=g\mu_{\text{B}}S\left(
1-\cos\theta\right)  $ $=\mathcal{\rho}_{\text{0}}+\mathcal{\rho}_{\text{s}}$,
where the "superfluid" part $\mathcal{\rho}_{\text{s}}=-g\mu_{\text{B}}%
S\cos\theta$ is conjugated with the magnon time-even current carried by the
$\theta$-fluctuations. Then, we obtain the magnon current via the continuity
equation $\mathcal{N}_{t}+{J}_{z}^{z}=0$. Here, we compute the magnon
current density carried by the traveling soliton in the limit of
$q\rightarrow1$, where the analytical solution\ with the intrinsic boost
transformation \ is available based on Eqs.(\ref{phiAn}) and (\ref{explthet}).

The additional magnon density associated with the solutions given by
Eqs.(\ref{phiAn}) and (\ref{explthet}) is
\[
\delta {\mathcal N}=-\frac{S}{2\beta^{2}}\tilde{\varphi}_{t}.
\]
Taking the time derivative of $\delta {\mathcal N}$ and using the property $\tilde
{\varphi}_{t}=-\upsilon_{0}\,\tilde{\varphi}_{z}$ we get $\delta
{\mathcal N}_{t}=-{J}_{z}^{z}$, where the current ${J}^{z}=S\upsilon
_{0}\theta_{1}$ has the explicit form
\begin{align}
{J}^{z}(z,t)  &  =S\sqrt{\frac{b}{2\beta^{2}}}\frac{\left(
1-k^{2}\right)  }{k}\frac{1}{1+V^{2}}\nonumber\\
&  \times\frac{\cosh  z  }{\cosh^{2}\left\{  \dfrac{k^{2}-1}%
{4k}\left(  \dfrac{1+k}{1-k}z+\mathcal{M}_{0}-t\right)  \right\}  }.
\label{Jcur}%
\end{align}
Note a significant difference of the result with the magnon current due to the
translational motion of the\textit{ whole} kink crystal considered by some of
the authors recently.\cite{BKO1} In the latter case, the sliding motion of the
whole kink crystal excites massive spin wave excitations of the $\theta$-mode
above the traveling state that is responsible for a magnon density transport.
In the present case, the nontrivial soliton solution itself carries \ a
localized magnon density (the magnon "droplet") due to the intrinsic boost symmetry.

In Fig. 3(a), we depict the background topological charge $\mathcal{Q}%
=\partial_{z}\varphi$ associated with the standing kink around $z=0$ in the
$q\rightarrow1$ limit, i.e., $\varphi(z)=2\sin^{-1}\left(  \text{tanh
}z\right)  $. In Figs. 3(b-1)-(b-6), we show the magnon density distribution
${J}^{z}(z,t)$\ carried by the traveling kink at the time $t=-20,$
$-4,$ $-2,$ $0,$ $2,$ $4,$, respectively, where the soliton travels from left
to right. The traveling soliton collides with the standing kink at $t=0$. It
is clearly seen that the magnon density is largely amplified when the soliton
"surfs" over the standing kink.\begin{figure}[ptb]
\begin{center}
\includegraphics[width=3.0in]{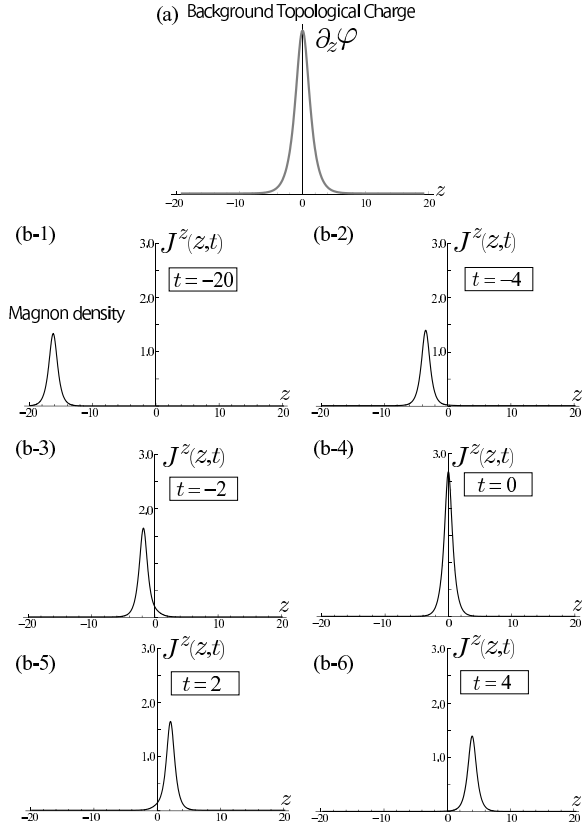}
\end{center}
\caption{(a) The background topological charge $\mathcal{Q}=\partial
_{z}\varphi$ associated with the standing kink. (b-1)-(b-6) The magnon
density distribution ${J}^{z}(z,t)$\ carried by the traveling kink at
the time $t=-20,$ $-4,$ $-2,$ $0,$ $2,$ $4,$ respectively. The soliton travels
from left to right. The magnon density is largely amplified when the soliton
"surfs" over the standing kink.}%
\end{figure}

\section{Concluding remarks}

In summary, by using the B\"{a}cklund transformation technique we investigated
soliton excitations in the chiral helimagnetic structure with the
antisymmetric Dzyaloshinskii-Moryia exchange and with the strong easy-plane
anisotropy, which is experienced by the external magnetic field applied
perpendicular to the modulation axis. The soliton we found was obtained as an
output of the BT from the kink crystal solution as an input. An essential
point is that\textit{ the traveling soliton cannot exist without the kink
crystal (soliton lattice) as a topological background configuration}. We may
say that the nontrivial topological object is excited over the topological
vacuum. The standing kink crystal enables the new soliton to emerge and
transport the magnon density. As compared with the motion of the whole kink
crystal with a heavy mass,\cite{BKO1} our new soliton is a well localized
object with a light mass. This object should be certainly, more easily
triggered off and propagate over the crystal.

We stress that our soliton has definite chirality, because of the presence of
the DM term $D\partial\varphi/\partial z$ in the Hamiltonian density
Eq.(\ref{Ham}). The presence of this term lifts the degeneracy between the
left-handed soliton and the right-handed antisoliton solutions. For example,
in Eq.(\ref{inspection}), the right-handed antisoliton  solutions may be given
by changing the sign of the phase gradient, i.e., $\varphi\left(  z\right)
=\pi-4\tan^{-1}\left[  F(z)\right]$, and $\tilde{\varphi}\left(  z,t\right)
=\pi-4\tan^{-1}\left[  V(z,t)\right]$. Although these solutions satisfy the
same SG equation as Eq. (\ref{SineGord}), their static energies are higher
than the left-handed soliton solution given by Eq.(\ref{inspection}).

Finally, it would be of interest to discuss a difference between the
conventional single Bloch wall and our soliton. The Bloch wall is formed
within a non-topological background, just a ferromagnet, and it dephases
easily. On the other hand, our soliton should be more robust against a
dephasing, because it emerges within the topological background (soliton
lattice). Chirality and topology support a stability of the moving soliton.
This consequence is quite obvious in the context of the soliton theory, but
may serve a quite new strategy in the field of spintronics. For example, in
the left-handed chiral crystal, only the left-handed kink crystal would be
formed and our soliton inherits with the corresponding chirality. Essentially,
the crystallographic chirality plays a role of protectorate for the background
chiral spin texture and causes the traveling soliton over the background.
This new traveling soliton can be regarded as a promising candidate to
transport magnetic information by using chiral helimagnet.

\begin{acknowledgments}
We acknowledge Yu.~A.~Izyumov for the interest to the work, and N.~E.~Kulagin pointed out us Ref.\cite{Kulagin}. J.~K. acknowledges
Grant-in-Aid for Scientific Research (A)(No.~18205023) and (C) (No.~19540371)
from the Ministry of Education, Culture, Sports, Science and Technology, Japan.
\end{acknowledgments}

\appendix

\section{Formation of the kink crystal state}

We here discuss the kink crystal (soliton lattice) formation from a general
view point. Let us consider a magnetic system described by the two-component
order parameter (OP) ($\eta$, $\xi$) with the Ginzburg-Landau
functional,\cite{Izyumov}
\begin{gather}
\Phi=\frac{1}{L}\int dz\left[  r(\eta\xi)+u(\eta\xi)^{2}+w(\eta^{n}+\xi
^{n})\right. \nonumber\\
\left.  +i\sigma\left(  \eta\frac{d\xi}{dz}-\xi\frac{d\eta}{dz}\right)
+\gamma\frac{d\eta}{dz}\frac{d\xi}{dz}\right]  , \label{6_1}%
\end{gather}
where the condition $u>0$, $\gamma>0$ ensures a stability of extremal points
of the functional. The signs of $w$ and $\sigma$ are arbitrary. We assume the
spin arrangement is uniform in the $x$ and $y$ directions and hence the volume
integral is implicitly reduced to one-dimensional integration over $z$-axis,
where $L$ is a crystal size in this direction. In the approximation of
constant OP modulus, $\rho=\text{const}$, when $\eta=\rho e^{i\varphi}$ and
$\xi=\rho e^{-i\varphi}$, the functional (\ref{6_1}) depends only on the phase
$\varphi,$%
\begin{align}
\Phi &  =r\rho^{2}+u\rho^{4}\nonumber\\
&  +\frac{1}{L}\int dz\left\{  \gamma\rho^{2}\left(  \frac{d\varphi}%
{dz}\right)  ^{2}+2\sigma\rho^{2}\frac{d\varphi}{dz}+2w\rho^{n}\cos
(n\varphi)\right\}  , \label{6_3}%
\end{align}
and includes $\rho$ as a parameter. Minimization of $\Phi$ with respect to
$\varphi$ results in the equation
\begin{equation}
\frac{d^{2}}{dz^{2}}(n\varphi)+v\sin(n\varphi)=0, \label{6_4}%
\end{equation}
where the effective anisotropy parameter is defined by $v=n^{2}(w/\gamma
)\rho^{(n-2)}$. The case of magnetic field corresponds to $n=1$.

Without of the nonlinear anisotropy term, Eq.(\ref{6_4}) is resolved by
$\varphi=Qz$ which describes an one-harmonic IC structure, for example, a
simple helimagnet, with the wave vector $Q=-\sigma/\gamma$. At finite $v,$ the
exact periodic solution is given by%
\begin{equation}
\sin\left[  \frac{n}{2}\varphi(z)\right]  =\text{sn}\left(  \frac{\sqrt{v}}%
{q}z,q\right)  . \label{6_7}%
\end{equation}
The elliptic modulus $q$ must be determined by minimizing the corresponding
energy,%
\begin{align}
\Phi_{\text{IC}}  &  =r\rho^{2}+u\rho^{4}-2\rho^{2}|\sigma|\frac{\pi\sqrt{v}%
}{nqK}\nonumber\\
&  +2\rho^{2}\gamma\frac{v}{n^{2}}\left(  \frac{q-2}{q^{2}}+\frac{4}{q^{2}%
}\frac{E}{K}\right)  , \label{6_8}%
\end{align}
where $K$ and $E$ denote the elliptic integrals of the first and second kind,
respectively. This procedure yields $q$ as a function of the anisotropy
parameter $v$,%
\begin{equation}
E/q=\sqrt{v_{c}/v}, \label{6_9}%
\end{equation}
with the critical anisotropy parameter being defined by $v_{c}=n^{2}\pi
^{2}\sigma^{2}/16\gamma^{2}.$ A change of $q$ from 0 to 1 corresponds to a
change of $v$ from 0 to $v_{c}$. Varying the parameter $v$ causes a drastic
change in the behavior of the amplitude (\ref{6_7}). The region of an almost
constant phase within the period $l$ comes up at $v\rightarrow v_{c}$, while
the phase rapidly changes at the ends of the period, where the overall phase
change is $2\pi/n$. The region of the constant phase increases as
$v\rightarrow v_{c}$. For $0<v<v_{c}$, the kink crystal phase is stabilized,
where a periodic array of C-phase regions separated by the kinks (solitons).
The spatial period is given by $l=4qK/\sqrt{v}$, and it diverges
logarithmically at $v\rightarrow v_{c}$, i.e. $q\rightarrow1$,%
\begin{equation}
l=(4q/\sqrt{v})\ln[4/\sqrt{1-q^{2}}]. \label{6_11}%
\end{equation}

\section{B\"{a}cklund transformation}

An existence of exact multi-soliton solutions is a peculiar property of the SG
equation, and the BT is a systematic way to obtain them. Indeed, let both
$\varphi_{0}$\ and $\varphi_{1}$\ are solutions of the SG equation
\[
\partial_{+}\partial_{-}\varphi=\sin\varphi
\]
written via light-cone coordinates $x^{+}=\left(  x+t\right)  /2$, and
$x^{-}=\left(  x-t\right)  /2$. Then, the B\"{a}cklund transformation
$\varphi_{1}=\mathcal{B}_{a}\left[  \varphi_{0}\right]  $\ is given by%
\begin{equation}
\partial_{\pm}\left(  \dfrac{\varphi_{1}\mp\varphi_{0}}{2}\right)
=e^{\pm\lambda}\sin\left(  \dfrac{\varphi_{1}\pm\varphi_{0}}{2}\right)  ,
\label{BTLC}%
\end{equation}
where $a=e^{\lambda}$ is called a scale parameter. The relation is consistent
with the SG equation, i.e., $\partial_{-}\partial_{+}\varphi_{0}=\sin
\varphi_{0}$\ and $\partial_{-}\partial_{+}\varphi_{1}=\sin\varphi_{1}$. Any
two functions $\varphi_{0}$\ and $\varphi_{1}$\ that satisfy the BT
necessarily solve the SG equation. Eq.(\ref{BTLC}) is nothing but Eq.
(\ref{BackTran}).

It turns out that analytical expression for multi-soliton solutions may be
outlined by an entirely algebraic procedure because the BT embodies a
nonlinear superposition principle known as Bianchi's permutation theorem.
Suppose that $\varphi_{0}$ is a seed SG solution, and $\varphi_{1,2}$ are the
BTs of $\varphi_{0}$, i.e.
\[
\varphi_{1}=\mathcal{B}_{a_{1}}\left[  \varphi_{0}\right]  , \quad\varphi
_{2}=\mathcal{B}_{a_{2}}\left[  \varphi_{0}\right]  .
\]
Two successive BTs commute, i.e. $\mathcal{B}_{a_{1}}\mathcal{B}_{a_{2}%
}=\mathcal{B}_{a_{2}}\mathcal{B}_{a_{1}}$, if the Bianchi's identity
\[
\varphi_{3}=\varphi_{0}+4\tan^{-1}\left[  \dfrac{a_{2}+a_{1}}{a_{2}-a_{1}}%
\tan\left(  \dfrac{\varphi_{2}-\varphi_{1}}{4}\right)  \right]
\]
is fulfilled. It means that the non-linear superposition rule holds
$\varphi_{3}=\mathcal{B}_{a2}\left[  \varphi_{1}\right]  $, $\varphi
_{3}=\mathcal{B}_{a1}\left[  \varphi_{2}\right]  $. This algebraic relation
indicates that a series of soliton solutions is given by $\varphi=4\tan
^{-1}\left[  f/g\right]  $, which supports the forms of Eq.(\ref{inspection}).

\section{Computation of the product $A(z)B(z)$}

Our objective is to find the product $A(z)B(z)$. By using Eqs.(\ref{cfA}%
,\ref{cfB}) we obtain
\begin{gather}
A(z)B(z)\nonumber\\
=\frac{1}{16k^{2}(1+F^{2})^{2}}\nonumber\\
\times\left[  16k^{2}F_{z}^{2}-F^{4}(k^{2}-1)^{2}-(k^{2}-1)^{2}-2F^{2}%
(1+k^{4}+6k^{2})\right] \nonumber\\
=\frac{1}{16k^{2}(1+F^{2})^{2}}\nonumber\\
\times\left[  16k^{2}F_{z}^{2}-4F^{2}(k^{2}+1)^{2}-\left(  F^{2}-1\right)
^{2}\left(  k^{2}-1\right)  ^{2}\right]  . \label{ABp}%
\end{gather}
The function $F(z)$ has the derivative
\[
F_{z}(z)=-\frac{1}{q}\frac{\text{dn}(z,q)}{1+\text{sn}(z,q)},
\]
that yields for the numerator in Eq.(\ref{ABp})
\begin{gather}
16k^{2}F_{z}^{2}-4F^{2}(k^{2}+1)^{2}-\left(  F^{2}-1\right)  ^{2}\left(
k^{2}-1\right)  ^{2}\nonumber\\
=16\frac{k^{2}}{q^{2}}\frac{\text{dn}^{2}(z,q)}{\left(  1+\text{sn}%
(z,q)\right)  ^{2}}-4(k^{2}+1)^{2}\frac{\text{cn}^{2}(z,q)}{\left(
1+\text{sn}(z,q)\right)  ^{2}}\nonumber\\
-4(k^{2}-1)^{2}\frac{\text{sn}^{2}(z,q)}{\left(  1+\text{sn}(z,q)\right)
^{2}}\nonumber\\
=\frac{4}{\left(  1+\text{sn}(z,q)\right)  ^{2}}\left(  \frac{4k^{2}}{q^{2}%
}-k^{4}-1-2k^{2}\right)  .
\end{gather}
Thus, we reproduce the result (\ref{AB}). Note that the product $A(z)B(z)$
embodies no coordinate dependence, hence it equals a constant $s$.

\section{Derivation of Eq.(\ref{SMPLB2})}

To derive the determining equation for the function $C_{1}(z)$ we rewrite
Eq.(\ref{Zdep}) through the function $U\left(  z,t\right)  $
\begin{gather*}
U_{z}=\frac{FF_{z}(k^{2}-1)}{Fk^{2}+2F_{z}k+F}\\
-\frac{(k^{2}-1)(F^{2}-1)\left(  F_{z}k^{2}+2F_{zz}k+F_z \right)  }{2\left(
Fk^{2}+2F_zk+F\right)  ^{2}}\\
+\frac{F^{2}(k^{2}+1)}{2k(1+F^{2})}\left(  U-\frac{(k^{2}-1)(F^{2}%
-1)}{2\left(  Fk^{2}+2kF_{z}+F\right)  }\right)  \\
-\frac{(k^{2}+1)}{2k(1+F^{2})}\left(  U-\frac{(k^{2}-1)(F^{2}-1)}{2\left(
Fk^{2}+2kF_{z}+F\right)  }\right)  \\
+\frac{F(k^{2}-1)}{2k(1+F^{2})}\left[  \left(  U-\frac{(k^{2}-1)(F^{2}%
-1)}{2\left(  Fk^{2}+2kF_{z}+F\right)  }\right)  ^{2}-1\right]  ,
\end{gather*}
and use Eq.(\ref{Ut}) to obtain the relation,
\begin{gather}
\frac{\mathcal{S}}{A}\left[  \frac{F^{2}(k^{2}+1)}{2k(1+F^{2})}+\frac{A_{z}%
}{A}-\frac{(k^{2}+1)}{2k(1+F^{2})}\right.  \nonumber\\
\left.  -\frac{F(k^{2}-1)}{2k(1+F^{2})}\frac{(k^{2}-1)(F^{2}-1)}{\left(
Fk^{2}+2kF_{z}+F\right)  }\right]  \tanh(T)\nonumber\\
+\left(  \frac{\mathcal{S}}{A}\right)  ^{2}\frac{1}{\cosh^{2}\left(  T\right)
}\nonumber\\
\times\left[  \frac{A_{z}}{A}\left(  At-C_{1}\right)  -\left(  A_{z}%
t-C_{1z}\right)  -\frac{F(k^{2}-1)}{2k(1+F^{2})}\right]  \nonumber\\
-\frac{(k^{2}+1)(k^{2}-1)(F^{2}-1)^{2}}{4k(1+F^{2})\left(  Fk^{2}%
+2kF_{z}+F\right)  }\nonumber\\
+\frac{(k^{2}-1)FF_{z}}{Fk^{2}+2kF_{z}+F}-\frac{(k^{2}-1)F}{2k(1+F^{2}%
)}\nonumber\\
+\left(  \frac{\mathcal{S}}{A}\right)  ^{2}\frac{(k^{2}-1)F}{2k(1+F^{2}%
)}\nonumber\\
-\frac{(k^{2}-1)(F^{2}-1)\left(  F_{z}k^{2}+2kF_{zz}+F_z\right)  }{2\left(
Fk^{2}+2kF_{z}+F\right)  ^{2}}\nonumber\\
+\frac{(k^{2}-1)^{3}F(F^{2}-1)^{2}}{8k(F^{2}+1)\left(  Fk^{2}+2kF_{z}%
+F\right)  ^{2}}\nonumber\\
=0,\label{BT2}%
\end{gather}
where $T=\left(  \mathcal{S}/A\right)  \left(  At-C_{1}\right)  $. Being
constant at any time moment, Eq.(\ref{BT2}) means that the coefficients before
$\tanh(T)$ and $\cosh^{-2}\left(  T\right)  $ turn simultaneously into zero.
With some tedious but a straightforward algebra (see below) one finds that
only the factor before $\cosh^{-2}\left(  T\right)  $ produces the non-trivial
result (\ref{SMPLB2}).

Indeed, consider the coefficient before $\tanh(T)$ in Eq.(\ref{BT2})
\begin{align}
&  \frac{\mathcal{S}}{A}\left[  \frac{F^{2}(k^{2}+1)}{2k(1+F^{2})}+\frac
{A_{z}}{A}-\frac{(k^{2}+1)}{2k(1+F^{2})}\right. \nonumber\\
&  \left.  -\frac{F(k^{2}-1)}{2k(1+F^{2})}\frac{(k^{2}-1)(F^{2}-1)}{\left(
Fk^{2}+2kF_{z}+F\right)  }\right] \nonumber\\
&  =\frac{\mathcal{S}}{2kA^{2}(1+F^{2})\left(  Fk^{2}+2kF_{z}+F\right)
}\nonumber\\
&  \times\left[  F(k^{2}-1)^{2}(F^{2}-1)A\right. \nonumber\\
&  -2kA_{z}(1+F^{2})\left(  F\left(  k^{2}+1\right)  +2kF_{z}\right)
\nonumber\\
&  \left.  +A(k^{2}+1)(1-F^{2})\left(  F\left(  k^{2}+1\right)  +2kF_{z}%
\right)  \right] \nonumber\\
&  =\frac{\mathcal{S}}{A^{2}(1+F^{2})\left(  Fk^{2}+2kF_{z}+F\right)
}\nonumber\\
&  \times\left[  A(1-F^{2})\left(  2kF+F_{z}(k^{2}+1)\right)  \right.
\nonumber\\
&  \left.  -A_{z}(1+F^{2})\left(  F\left(  k^{2}+1\right)  +2kF_{z}\right)
\right]  . \label{tanhT}%
\end{align}
Now we use Eq.(\ref{cfA}) to find
\begin{gather}
\frac{A_{z}}{A}=\frac{1}{1+F^{2}}\frac{1}{F\left(  k^{2}+1\right)  +2kF_{z}%
}\nonumber\\
\times\left[  2kF_{zz}(1+F^{2})+\left(  k^{2}+1\right)  F_{z}(1+F^{2})\right.
\nonumber\\
\left.  -4kFF_{z}^{2}-2F^{2}F_{z}\left(  k^{2}+1\right)  \right]  .
\label{ratioA}%
\end{gather}
The function $F(z)$ obeys the equation
\[
(1+F^{2})F_{zz}-2FF_{z}^{2}=-F(F^{2}-1),
\]
which is derived from the SG equation (\ref{SineGord}). Together with
Eq.(\ref{ratioA}) this produces
\[
\frac{A_{z}}{A}=\frac{1-F^{2}}{1+F^{2}}\frac{F_{z}\left(  k^{2}+1\right)
+2kF}{F\left(  k^{2}+1\right)  +2kF_{z}}.
\]
Plugging this into the right-hand side of Eq.(\ref{tanhT}) we come to zero.

After lengthy manipulation it may be shown that the free terms in
Eq.(\ref{BT2}), i.e. those without either $\tanh(T)$ or $\cosh^{-2}\left(
T\right)  $ factors, are simplified to
\begin{align*}
&  -\left(  k^{2}-1\right)  \left[  q^{2}k^{4}-2k^{2}q^{2}(8\mathcal{S}%
^{2}-1)-4k^{2}+q^{2}\right] \\
&  \times\text{cn}\left(  z,q\right)  \,\frac{\left[  \left(  k^{2}+1\right)
q\,\text{cn}\left(  z,q\right)  -2k\,\text{dn}\left(  z,q\right)  \right]
^{3}}{16k^{3}q^{5}\left[  1+\text{sn}\left(  z,q\right)  \right]  ^{3}}.
\end{align*}
The factor $q^{2}k^{4}-2k^{2}q^{2}(8\mathcal{S}^{2}-1)-4k^{2}+q^{2}$ equals to
zero bearing in mind Eq.(\ref{AB}).

\section{Derivation of Eq.(\ref{Mzsol})}

To obtain $\mathcal{M}(z),$ the derivative (\ref{Mzder}) is split into two
parts,%
\[
\mathcal{M}_{z}(z)=\mathcal{M}_{0}(z)+\mathcal{M}_{1}(z),
\]
where
\begin{align*}
\mathcal{M}_{0}(z)  &  =\frac{(1-k^{4})q^{2}\text{cn}^{2}(z,q)}{(1+k^{2}%
)^{2}q^{2}-4k^{2}-(1-k^{2})^{2}q^{2}\text{sn}^{2}(z,q)},\\
\mathcal{M}_{1}(z)  &  =\frac{2kq(1-k^{2})\text{cn}(z,q)\text{dn}%
(z,q)}{(1+k^{2})^{2}q^{2}-4k^{2}-(1-k^{2})^{2}q^{2}\text{sn}^{2}(z,q)},
\end{align*}
and both terms are separately considered.

The integration of $\mathcal{M}_{0}(z)$ is straightforwardly performed and one
obtains
\begin{align*}
&  \int dz\,\mathcal{M}_{0}(z)\\
&  =-\frac{k^{2}+1}{k^{2}-1}z+\frac{4k^{2}(1-q^{2})(1+k^{2})}{(1-k^{2}%
)[(1+k^{2})^{2}q^{2}-4k^{2}]}\\
&  \times\Pi\left(  \frac{(1-k^{2})^{2}q^{2}}{(1+k^{2})^{2}q^{2}-4k^{2}%
},\text{am}(z,q),q^{2}\right)  ,
\end{align*}
where%
\[
\Pi(u,a,q^{2})=\int_{0}^{u}\frac{q^{2}\text{sn}(a,q)\text{cn}(a,q)\text{dn}%
(a,q)\text{sn}^{2}(u,q)}{1-q^{2}\text{sn}^{2}(a,q)\text{sn}^{2}(u,q)}\,du
\]
is the elliptical integral of the third kind.

Taking into account the relationships $(1+k^{2})^{2}q^{2}-4k^{2}=16q^{2}%
k^{2}{\mathcal{S}}^{2}$ and $\text{cn}(z,q)\text{dn}(z,q)=({d}/{dz}%
){\text{sn}(z,q)}$ the second term can be written as follows
\begin{align*}
\mathcal{M}_{1}(z)  &  =\frac{(1-k^{2})}{4{\mathcal{S}}}\left[  \frac
{1}{4{\mathcal{S}}qk-(1-k^{2})q\text{sn}(z,q)}\right. \\
&  \left.  +\frac{1}{4{\mathcal{S}}qk+(1-k^{2})q\text{sn}(z,q)}\right]
\frac{d}{dz}\text{sn}(z,q)
\end{align*}
that yields the desired result
\[
\int dz\,\mathcal{M}_{1}(z)=\frac{1}{4q{\mathcal{S}}}\log\left\vert
\frac{4{\mathcal{S}}k-(k^{2}-1)\,\text{sn}(z,q)}{4{\mathcal{S}}k+(k^{2}%
-1)\,\text{sn}(z,q)}\right\vert .
\]

\end{document}